\documentclass[11pt,a4paper]{article}
\newcommand{\be}{\begin{equation}}
\newcommand{\ee}{\end{equation}}
\def\mz{{\mbox{\tiny{(0)}}}}
\def\mt{{\mbox{\tiny{(1)}}}}
\def\my{{\mbox{\tiny{(2)}}}}
\newcommand{\Gammabol}{\Gamma{}}
\newcommand{\Rbol}{R{}}
\newcommand{\nabol}{\nabla{}}
\newcommand{\onehalf}{{\textstyle{\frac{1}{2}}}}

\usepackage{revsymb}
\usepackage{bm}
\usepackage{amssymb}
\usepackage{amsmath}
\usepackage{graphicx}
\usepackage{color}
\RequirePackage[numbers,sort&compress]{natbib}

\begin{document}

\noindent
{\Large \bf Gravitational waves: a foundational review}\footnote{This monograph is a compilation of the results previously published in Refs.~\cite{waves} and \cite{gw2}.}
\vskip 0.7cm
\noindent
{\bf J. G. Pereira}
\vskip 0.1cm \noindent
{\it Instituto de F\'{\i}sica Te\'orica,
Universidade Estadual Paulista \\
Caixa Postal 70532-2, 01156-970 S\~ao Paulo, SP, Brazil}\\
E-mail: jpereira@ift.unesp.br

\vskip 0.8cm

\begin{quote}
{\bf Abstract.}~{\footnotesize The standard approach to the gravitational waves theory is critically reviewed. Contrary to the prevalent understanding, it is pointed out that this theory contains many conceptual problems and technical obscure points that require further analysis.}

\end{quote}

\section{Introduction}

For many years in the past, the existence of gravitational waves was a controversial issue. The discovery of a binary pulsar whose orbital period changes in accordance with the predicted gravitational wave emission~\cite{HT} put an end to that controversy. In fact, that discovery provided a compelling evidence for the existence of gravitational waves (for a textbook reference, see Ref.~\cite{maggiore}). That evidence, however, did not provide any clue on their form and effects. The only it has done was to confirm the quadrupole radiation formula. Despite this fact, together with the quadrupole radiation formula, the standard linear approach to the gravitational waves theory became widely considered a finished topic, a theory not to be questioned anymore (see, for example, Ref.~\cite{giants}, page 313). In other words, it became a dogma.

As a careful analysis of the current gravitational wave theory shows, however, it is actually plagued by many obscure points \cite{cs1}. From one hand, owing to the nonlinear nature of gravitation, which makes it difficult to deal with, the existence of some obscure and controversial points is fully understandable. On the other hand, these difficulties cannot be used as an excuse for our leniency with the established theory. In these notes, even at the risk of committing a heresy, I will critically review the foundations of the standard linear approach to the gravitational waves theory, and will point out precisely where it lacks consistency and why it requires further attention.

\section{The meaning of being linear}
\label{non}

It is instructive to begin by studying some properties of linear waves. We begin by analysing gauge field waves---of which the electromagnetic wave is an example---and then we consider gravitational waves. Similarities and mainly the differences between them are discussed.

\subsection{Gauge field waves}

It is well-known that, in order to transport its own charge (or source), a gauge field must satisfy a nonlinear field equation. For example, the field equations of chromodynamics must be nonlinear to allow the gauge field to transport color charge. In the language of differential forms, the Yang-Mills equation\footnote{The Yang-Mills theory will be adopted as the paradigm of nonlinear gauge theories.} is written as \cite{itzu}
\be
dF - j = J,
\label{ge}
\ee
where $F = DA$ is the field strength of the gauge potential $A$, with $D$ the gauge covariant derivative. In addition, $j$ stands for the gauge pseudo-current and $J$ is the source current. Due to the property
\be
d d = 0,
\label{Poincare}
\ee
known as Poincar\'e lemma \cite{livro}, the field equation implies the conservation of the total current:
\be
d \left(j + J \right) = 0.
\ee

Electromagnetism is a particular case of Yang-Mills theories, with the Abelian unitary group $U(1)$ as the gauge group. In this case, the Yang-Mills equation reduces to the {\em linear} Maxwell equation
\be
dF = J,
\label{Me}
\ee
where $F = dA$ the electromagnetic field strength, with $A$ the electromagnetic potential. The source $J$ in this case is the electric current, which is conserved on account of the Poincar\'e lemma:
\be
d J = 0.
\label{EmSouCon}
\ee
This conservation law says that a source cannot lose electric charge when emitting electromagnetic waves. In fact, remembering that currents are quadratic in the field variable, {\em the linearity of Maxwell equation restricts the gauge self-current $j$ to be linear, and consequently to vanish}
\be
j = 0.
\ee
This is the reason why an electromagnetic wave is unable to transport its own source, that is, electric charge, a result consistent with the source conservation law~(\ref{EmSouCon}). Observe that the source current $J$ is quadratic in the source field variables, but linear in the electromagnetic field. Differently from the self-current $j$, therefore, the linearity of Maxwell equation does not restrict it to vanish. Of course, {\em since neither energy nor momentum is source of electromagnetic field, the energy-momentum current does not appear explicitly in the electromagnetic field equation, and for this reason the linearity of Maxwell equation do not restrict the energy-momentum tensor of the electromagnetic field to be linear. This means that, even though the linearity of the electromagnetic waves preclude them to carry electric charge, they can carry energy and momentum, with the amount transported given by the (quadratic) Poynting vector.}

\subsection{Gravitational waves}

As is well-known, there is no an {\em invariant} lagrangian for general relativity that depends on the tetrad and its first derivatives only. What exists is the second-order Einstein-Hilbert invariant lagrangian ${\mathcal L}_{EH}$, in which the second-derivative terms reduce to a total divergence. This lagrangian can be split in the form ($\kappa = 8 \pi G/c^4$)\footnote{We are going to use the first half of the Latin alphabet ($a, b, c, \dots$ = 0, 1, 2, 3) to denote algebraic (or tangent space) indices, and the Greek alphabet ($\alpha, \beta, \mu, \dots = 0, 1, 2, 3$) to denote spacetime indices. The second half of the Latin alphabet ($i, j, k, \dots = 1, 2, 3$) will be used to denote space indices.}
\be
{\mathcal L}_{EH} \equiv -\, \frac{e}{2 \kappa} \, R \, = \, {\mathcal L}_g +
\partial_\mu (e \, w^\mu),
\ee
where ${\mathcal L}_g$ is a (non-invariant) lagrangian that depends on the tetrad and its first derivatives only, $w^\mu$ is a four-vector, and $e = \det (e^a{}_\mu)$ with $e^a{}_\mu$ the tetrad field. Denoting by ${\mathcal L}_s$ the source Lagrangian, the Euler-Lagrange equation obtained from ${\mathcal L} = {\mathcal L}_g + {\mathcal L}_s$ is the potential (or lagrangian) form of Einstein equation \cite{Moller1}
\begin{equation}
d(eH_a) - \kappa \, e t_a = \kappa \, e T_a,
\label{dH}
\end{equation}
where
\be
H_a = - \frac{\kappa}{e} \frac{\partial {\mathcal L}_g}{\partial d e^a}
\ee
is the gravitational field excitation 2-form, also known as superpotential, and $e^a = e^a{}_\mu dx^\mu$. In addition,
\be
t_a = - \frac{1}{e} \frac{\partial {\mathcal L}_g}{\partial e^a}
\ee
stands for the gravitational self-current, which in this case represents the gravitational energy-momentum pseudotensor, and
\be
T_a = - \frac{1}{e} \frac{\partial {\mathcal L}_s}{\partial e^a} 
\ee
is the source energy-mo\-men\-tum current. Notice that in this form, Einstein equation~(\ref{dH}) is similar, in structure, to the Yang-Mills equation (\ref{ge}). {Its main property is to explicitly exhibit the complex defining the energy-momentum pseudo-current of the gravitational field}. From the Poincar\'e lemma~(\ref{Poincare}), the total energy-momentum density is found to be conserved as a consequence of the field equation:
\be
d [e(t_a + T_a)] = 0.
\ee

Now, considering that the sources of gravitational waves are at enormous distances from Earth, it is reasonable to assume that the amplitude of a gravitational wave when reaching a detector on Earth will be very small. This allows the use of a perturbative analysis where the gravitational field, or tetrad, is expanded in the form
\be
e^a = \delta^a + \varepsilon \, e_{\mt}^ a + \varepsilon^2 e_{\my}^a + \cdots ,
\label{PotExpan}
\ee
where $\delta^a$ is a trivial tetrad related to Minkowski spacetime, and $\varepsilon$ is a small parameter introduced to label the successive orders of the perturbation scheme. In the linear, or first-order approximation, the gravitational field equation becomes mathematically similar to Maxwell equation. In fact, at this order the field equation~(\ref{dH}) reduces to
\begin{equation}
dH^\mt_a = \kappa \, T^\mt_a
\label{dH1}
\end{equation}
which is mathematically similar to the Maxwell equation (\ref{Me}).

However, in spite of the similarity, there is a fundamental difference between the two cases. As we have already discussed, the linearity of Maxwell equation restricts the electromagnetic self-current to vanish: $j = 0$. As a consequence, electromagnetic waves are unable to transport their own source, that is, electric charge. In the same token, since the gravitational energy-momentum pseudo-current $t_a$ is at least quadratic in the field variables \cite{living}, it vanishes in the linear approximation:
\be
t^\mt_a = 0
\label{t1}
\ee
leading then to the linear field equation (\ref{dH1}). This means that {\em linear gravitational waves are unable to transport their own source, that is, energy and momentum}. This property is consistent with the Poincar\'e lemma, which when applied to the linear field equation~(\ref{dH1}) implies that the first-order source energy-momentum tensor is conserved:
\be
d T^\mt_a = 0.
\label{lcl}
\ee
Strictly speaking, this conservation law says that, at this order, a mechanical system cannot lose energy and momentum. It is usual in gravitational wave theory to argue that, {\em if a (linear) electromagnetic wave is able to transport energy and momentum, a linear gravitational wave might also be able to do it. In the light of the above discussion on the field ability to transport it own source, however, this argument is easily seen to be conceptually misleading.}

\section{The standard approach to gravitational waves}

\subsection{Linear gravitational waves}
\label{TTcs}

The {\em invariant formalism} of differential forms is useful for theoretical discussions. When talking about experiments, however, which are always performed in a particular frame and using apparatuses that presuppose a specific coordinate system, the use of a {\em covariant formalism} in terms of components is mandatory. Thus, in consonance with the tetrad expansion (\ref{PotExpan}), the metric tensor $g_{\mu \nu} = e^a{}_\mu e_{a \nu}$ is expanded in the form
\be
g_{\mu \nu} = \eta_{\mu \nu} + \varepsilon \, h_{\mt \mu \nu} +
\varepsilon^2 h_{\my \mu \nu} + \cdots 
\label{MetExpan}
\ee
with $\eta_{\mu \nu}$ the metric of the background Minkowski spacetime. A quite convenient class of coordinates to study waves is the class of harmonic coordinates, which is defined by the condition
\be
g^{\mu \nu} \, \Gamma^\rho{}_{\mu \nu} = 0.
\label{HarmoCoor}
\ee
At the first order it becomes
\be
\partial_\mu h^{\,\mu}_{\mt \nu} = \onehalf \partial_\nu h_{\mt},
\label{hcc}
\ee
where we used the notation $h_{\mt} = h^{\,\mu}_{\mt \mu}$. In these coordinates, the sourceless version of the linear field equation (\ref{dH1}) assumes the form
\be
\Box \, h^{\,\mu}_{\mt \nu} = 0
\label{we1}
\ee
with $\Box$ the flat spacetime d'Alambertian operator.
A monochromatic plane-wave solution to this equation has the form
\be
h^{\,\mu}_{\mt \nu} = A^{\,\mu}_{\mt \nu}  \exp(i k_{\rho} x^\rho)
\label{pw}
\ee
where $A^{\,\mu}_{\mt \nu}$ is the (symmetric) polarization tensor and $k^\rho$ is the wave vector, which on account of the field equation~(\ref{we1}) is found to satisfy the dispersion relation
\be
k_{\rho}  k_{}^\rho = 0.
\label{12}
\ee
In this case, the harmonic coordinate condition (\ref{hcc}) becomes
\be
k_\mu  h^{\,\mu}_{\mt \nu} = \onehalf k_\nu h_{\mt}.
\label{13}
\ee

Analogously to the gauge choice in electromagnetism, it is possible to choose, within the class of harmonic coordinates, one specific coordinate system. Once this is done, the coordinate system becomes completely specified, and the remaining components of $h^{\,\mu}_{\mt \nu}$ turn out to represent only physical degrees of freedom. A quite convenient choice is the so-called {\it transverse-traceless} coordinates (sometimes called transverse-traceless {\em gauge}), in which
\be
h_{\mt} \equiv h^{\,\mu}_{\mt \mu} = 0 \qquad \mbox{and} \qquad
h^{\,\mu}_{\mt \nu} U_{\mbox{\tiny{(0)}}}^\nu = 0
\label{TTA}
\ee
with $U_{\mbox{\tiny{(0)}}}^\nu$ an arbitrary, constant four-velocity. In these coordinates, the harmonic coordinate condition (\ref{13}) assumes the form
\be
k_\mu  h^{\,\mu}_{\mt \nu} = 0.
\label{13bis}
\ee

Although the coordinate system has been completely specified, there is still the freedom to choose different Lorentz frames $e^a$. In particular, it is always possible to choose a frame in which the constant four-velocity $U_{\mbox{\tiny{(0)}}}^\mu$ acquires the form (see Section~\ref{foGDE} below)
\be
U_{\mbox{\tiny{(0)}}}^\mu \equiv \delta^\mu_0 = (1, 0, 0, 0).
\label{U00}
\ee
In this case, as can be seen from the second of the Eqs.~(\ref{TTA}),
\be
h^{\,\mu}_{\mt 0} = 0
\ee
for all $\mu$. Orienting the frame in such a way that the wave travels in the $z$ direction, the wave vector assumes the form
\be
k^\rho = \left({\omega}/{c}, 0, 0, {\omega}/{c} \right),
\label{kzdir}
\ee
and the physical components of the wave are found to be
\be
h^{\,x}_{\mt x} = - h^{\, y}_{\mt y} \qquad \mbox{and} \qquad
h^{\,x}_{\mt y} = h_{\mt y}{}^{x}.
\label{FOmetric}
\ee
Linear waves satisfying these conditions are said to represent a plane gravitational wave in transverse-traceless coordinates.\footnote{It is important to note that, since the metric $g_{\mu \nu} = \eta_{ab} \, e^a{}_\mu e^b{}_\nu$ is invariant under changes of frames, the metric perturbation $h^{\,\mu}_{\mt \nu}$ will also be invariant. As a consequence, the Levi-Civita connection, as well as the corresponding Riemann tensor, will of course be frame invariant.}

\subsection{The first-order geodesic deviation equation}
\label{foGDE}

Let us consider two nearby particles separated by the four-vector $\xi^\alpha$. This vector obeys the geodesic deviation equation
\be
\nabol_U \, \nabol_U \xi^\alpha = \Rbol^\alpha{}_{\mu \nu \beta} \,
U^\mu \, U^\nu \, \xi^\beta,
\label{gde0}
\ee
where $U^\mu = dx^\mu/ds$ is the four-velocity of the particles, with $ds = (g_{\mu \nu} \, dx^\mu dx^\nu)^{1/2}$. On account of the metric expansion (\ref{MetExpan}), each order of the correspondding Riemann tensor expansion
\be
\Rbol^\alpha{}_{\mu \nu \beta} = \varepsilon \, \Rbol_{\mbox{\tiny{(1)}}}^\alpha{}_{\mu \nu \beta} + \varepsilon^2 \, \Rbol_{\mbox{\tiny{(2)}}}^\alpha{}_{\mu \nu \beta} + \dots 
\ee
will give rise to a different contribution to $\xi^\alpha$. For consistence reasons, therefore, this vector must also be expanded,
\be
\xi^\alpha = \xi_{\mbox{\tiny{(0)}}}^\alpha + \varepsilon \, \xi_{\mbox{\tiny{(1)}}}^\alpha +
\varepsilon^2 \, \xi_{\mbox{\tiny{(2)}}}^\alpha + \dots \; ,
\ee
where $\xi_{\mbox{\tiny{(0)}}}^\alpha$ represents the initial, that is, undisturbed separation between the two particles. Of course, as the four-velocity $U^\mu$ depends on the gravitational field, it must also be expanded,
\be
U^\mu = U_{\mbox{\tiny{(0)}}}^\mu + \varepsilon \, U_{\mbox{\tiny{(1)}}}^\mu +
\varepsilon^2 \, U_{\mbox{\tiny{(2)}}}^\mu + \dots
\ee
with $U_{\mbox{\tiny{(0)}}}^\mu$ a constant arbitrary four-velocity, which depends on the frame from which the phenomenon will be observed and measured. Choosing a frame fixed at one of the particles, usually called {\it proper frame}, the four-velocity $U_{\mbox{\tiny{(0)}}}^\mu$ can be expressed in terms of the observer proper time $s_{\mbox{\tiny{(0)}}}$ in the form $U_{\mbox{\tiny{(0)}}}^\mu = dx^\mu/ds_{\mbox{\tiny{(0)}}}$, where
\be
ds_{\mbox{\tiny{(0)}}}^2 = \eta_{\mu \nu} \, dx^\mu dx^\nu
\label{ds0}
\ee
is the flat spacetime quadratic interval. Since in the proper frame the particles are initially at rest, we have
\be
U_{\mbox{\tiny{(0)}}}^\mu \equiv \delta^\mu_0 = (1, 0, 0, 0),
\label{U0}
\ee
which means that, in this frame, the proper time $s_{\mbox{\tiny{(0)}}}$ coincides with the coordinate $x^0$.

At first order, the geodesic deviation equation~(\ref{gde0}) assumes the form
\be
\frac{d^2 \xi_{\mbox{\tiny{(1)}}}^\alpha}{ds_{\mbox{\tiny{(0)}}}^2} +
U^\rho_{\mbox{\tiny{(0)}}} \, \partial_\rho \left(\Gammabol_{{\mbox{\tiny{(1)}}}}^\alpha{}_{\beta \gamma} \, U^\gamma_{\mbox{\tiny{(0)}}} \right) \xi_{\mbox{\tiny{(0)}}}^\beta  =
\Rbol_{\mbox{\tiny{(1)}}}^\alpha{}_{\mu \nu \beta} \, U^\mu_{\mbox{\tiny{(0)}}} \, U^\nu_{\mbox{\tiny{(0)}}} \; \xi_{\mbox{\tiny{(0)}}}^\beta
\label{gde01}
\ee
with
\be
\Gammabol_{{\mbox{\tiny{(1)}}}}^\alpha{}_{\beta \gamma} = 
\onehalf \, \eta^{\alpha \rho} \left(\partial_\beta h_{\mt\rho \gamma} +
\partial_\gamma h_{\mt\rho \beta} - \partial_\rho h_{\mt \beta \gamma} \right)
\label{ChristoConne}
\ee
the first-order Christoffel connection. Substituting $U^\mu_{\mbox{\tiny{(0)}}}$ given by Eq.~(\ref{U0}), it reduces to
\be
\frac{d^2 \xi_{\mbox{\tiny{(1)}}}^\alpha}{ds_{\mbox{\tiny{(0)}}}^2} + \partial_0 \Gammabol_{{\mbox{\tiny{(1)}}}}^\alpha{}_{\beta 0} \; \xi_{\mbox{\tiny{(0)}}}^\beta  = \Rbol_{\mbox{\tiny{(1)}}}^\alpha{}_{00\beta} \; \xi_{\mbox{\tiny{(0)}}}^\beta.
\label{gde2}
\ee
Using then the first-order Riemann tensor
\be
\Rbol_{\mbox{\tiny{(1)}}}^\alpha{}_{\mu \nu \beta} =
\partial_\nu \Gammabol_{\mt}^\alpha{}_{\mu \beta} -
\partial_\beta \Gammabol_{{\mbox{\tiny{(1)}}}}^\alpha{}_{\mu \nu},
\label{R1}
\ee
it becomes
\be
\frac{d^2 \xi_{\mbox{\tiny{(1)}}}^\alpha}{ds_{\mbox{\tiny{(0)}}}^2} + \partial_0 \Gammabol_{{\mbox{\tiny{(1)}}}}^\alpha{}_{\beta 0} \; \xi_{\mbox{\tiny{(0)}}}^\beta = \left( \partial_0 \Gammabol_{{\mbox{\tiny{(1)}}}}^\alpha{}_{\beta 0} -
\partial_\beta \Gammabol_{{\mbox{\tiny{(1)}}}}^\alpha{}_{0 0} \right) \xi_{\mbox{\tiny{(0)}}}^\beta.
\label{gde2bis}
\ee
This is the first-order geodesic deviation equation as seen from the proper frame.

\subsection{Effects on free particles}
\label{standardEFP}

Although the frame has already been chosen to be the proper frame, the coordinate system remains arbitrary up to this point. Using this arbitrariness, and motivated perhaps by the feeling that linear gravitational waves should produce some effects on free particles, rather than no effects, the standard approach adopts the following procedure. First, one elects a locally inertial coordinate system in which the Christoffel connection $\Gammabol_{{\mbox{\tiny{(1)}}}}^\alpha{}_{\beta \gamma}$ vanishes at the origin $x^i = 0$ of the proper frame for all $x^0$, which is the point where the computation is being performed. This means that $\partial_0 \Gammabol_{{\mbox{\tiny{(1)}}}}^\alpha{}_{\beta \gamma}$ also vanishes at that point \cite{mtw}. This result is then used to eliminate the connection-term from the left-hand side of Eq.~(\ref{gde2bis}), but not the very same connection term from the right-hand side, which yields 
\be
\frac{d^2 \xi_{\mbox{\tiny{(1)}}}^\alpha}{ds_{\mbox{\tiny{(0)}}}^2} = \left( \partial_0 \Gammabol_{{\mbox{\tiny{(1)}}}}^\alpha{}_{\beta 0} -
\partial_\beta \Gammabol_{{\mbox{\tiny{(1)}}}}^\alpha{}_{0 0} \right) \xi_{\mbox{\tiny{(0)}}}^\beta.
\label{gde2bisa}
\ee
This procedure is usually justified by arguing that the term inside the parentheses on the right-hand side is the first-order Riemann tensor. Since this tensor is {invariant} under general coordinate transformations, it can be computed in any coordinate system, and not necessarily in locally inertial coordinates. Taking advantage of this freedom, although the left-hand side was computed in locally inertial coordinates, one substitutes the Riemann tensor on the right-hand side written in transverse-traceless coordinates. In this case, $\partial_\beta \Gammabol_{{\mbox{\tiny{(1)}}}}^\alpha{}_{0 0} =0$, and the geodesic deviation equation reduces to
\be
\frac{d^2 \xi_{\mbox{\tiny{(1)}}}^\alpha}{ds_{\mbox{\tiny{(0)}}}^2} = \partial_0 \Gammabol_{{\mbox{\tiny{(1)}}}}^\alpha{}_{\beta 0} \, \xi_{\mbox{\tiny{(0)}}}^\beta.
\label{gde2bisb}
\ee
Expressing the right-hand side in terms of the metric perturbation, one gets
\be
\ddot \xi_{\mbox{\tiny{(1)}}}^\alpha = \onehalf h^{\,\alpha}_{\mt \beta} \, \xi_{\mbox{\tiny{(0)}}}^\beta.
\label{wrongGDE}
\ee
where we have denoted time derivative with a dot. This is the usual equation describing the effects of gravitational waves on free particles. It is ironic to observe that these effects come entirely from the connection term on the right-hand side of Eq.~(\ref{gde2bisb}), which is exactly the same term deliberately removed from the left-hand side of Eq.~(\ref{gde2bis}).

Let us consider now the first-order solution (\ref{FOmetric}). As is well known, there are two possible polarizations. For definiteness, we consider the polarization in which
\be
h^{\,x}_{\mt x} \neq 0 \qquad \mbox{and} \qquad
h^{\,x}_{\mt y} = 0.
\label{FOpola}
\ee
The effects of these waves can then be obtained from the geodesic deviation equation (\ref{wrongGDE}). Consider a ring of particles initially at rest in the $x-y$ plane. When a linear gravitational wave passes through those particles, it makes them to oscillate orthogonally around the original position. A ring of particles would be distorted in such a way that it would become an ellipse, first (let us say) horizontally, returning subsequently to ring, and then vertically as $h^{\,x}_{\mt x}$ changes sign, and so on. This behavior is depicted in Fig.~1 below.
\begin{figure}[ht]
\begin{center}
\scalebox{0.7}{\includegraphics{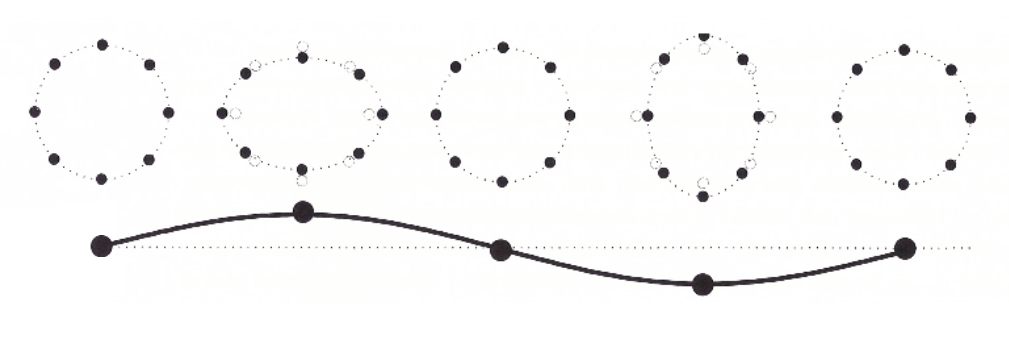}}
\label{fig1}
\vskip -0.8cm
\caption{\it Effects produced on a ring of free particles by a linear gravitational wave, according to the standard approach.}
\end{center}
\end{figure}

\section{Questionable and obscure points}

On account of the usual assumptions concerning the linearity of the gravitational waves, as discussed in the previous section, the standard linear approach to gravitational waves turns out to be plagued by many obscure points. In this section, some of these points are discussed. 

\subsection{Linear or nonlinear? That is the question}
\label{LinNonLin}

Even though there seems to be a general agreement that the transport of energy and momentum by gravitational waves is a nonlinear phenomenon (see, for example, Ref.~\cite{bondi}), {instead of going to the second order}, the standard approach to gravitational waves adopts a kind of ``mixed procedure'', which {\em consists basically in assuming that gravitational waves carry energy (are nonlinear), but at the same time, because the amount of energy transported is so small, it is also assumed that its dynamics can be approximately described by a linear equation} \cite{all}. More precisely, it can be described by the (sourceless version of the) linear wave equation (\ref{dH1}).

Conceptually speaking, however, this is a questionable assumption. The reason is that either a gravitational wave does or does not carry energy. If it carries, it cannot satisfy a linear equation. If applied to a Yang-Mills propagating field, it would correspond to assuming that, for a gauge field with small-enough amplitude, its evolution could be accurately described by a linear equation. Of course, this is plainly wrong: a Yang-Mills propagating field must be nonlinear to carry its own source, otherwise it is not a Yang-Mills field. Analogously, a gravitational wave must be nonlinear to transport its own source, otherwise it is not a gravitational wave. {\em This is not a matter of approximation, but a conceptual question.}\footnote{We remark that even the well-known {\em exact plane gravitational wave} solution of Einstein equations \cite{exact} transports neither energy nor momentum \cite{exact2}. This is in agreement with the nonlinear nature of the energy-momentum transport by gravitational waves.}

\subsection{Messing with the perturbation scheme}
\label{mess}

One of the main problems of the standard approach is the computation of the energy and momentum transported by linear gravitational waves. Since the first-order energy-momentum pseudo-current vanishes, the usual procedure is to make use of the second-order energy-momentum pseudo-current \cite{all}. The argument used to justify such procedure is that this is similar to the electromagnetic wave, which despite being linear, its energy-momentum tensor is quadratic in the field variables. However, as discussed in Section~\ref{non}, this is a misleading argument. The reason is that the linearity of Maxwell equation means that electromagnetic waves are unable to transport its own source, that is, electric charge. On the other hand, the linearity of the first-order gravitational field equation means that gravitational waves are unable to transport its own source, that is, energy and momentum. In fact, the linearity of the first-order gravitational field equation restricts the energy-momentum current to be {\em linear}, and consequently to vanish at this order. A quadratic (non-vanishing) energy-momentum pseudotensor can only appear in orders higher than one. This is the case, for example, of the second-order gravitational field equation~(\ref{dH2}) below, where $t^\my_a$ represents the second-order gravitational energy-momentum pseudotensor. {\em We arrive then at the surrealistic situation in which gravitational waves are assumed to be given by the solutions to the first-order gravitational field equation (\ref{dH1}), whilst the energy and momentum they transport are computed in terms of the energy-momentum current appearing in the second-order field equation~(\ref{dH2})}. It is obvious that this tensor represents the energy and momentum transported by second-order gravitational waves---and not by first-order waves, as usually assumed. The concomitant use of quantities belonging to different orders of the perturbation scheme constitutes a clear violation of the perturbation method itself, being unacceptable from any point of view.

\subsection{Tidal effects an the wave frequency}
\label{freque}

Gravitational waves are generated, and act on free particles through tidal effects (see Ref.~\cite{giants}, page 310, for a detailed explanation). These effects are well-known to be produced by inhomogeneities in the gravitational field, and like the ocean tides on Earth occur twice for each complete cycle of the system. In fact, according to the quadrupole radiation formula, gravitational radiation comes out from the source with a frequency that is twice the source frequency \cite{schutz}. Now, for symmetry reasons, the kind motion by which matter 
produces gravitational waves is also the action a gravitational wave will have on matter. This means that a gravitational wave of frequency $2 \omega$ should produce on free particles an oscillatory motion with frequency $\omega$. Although this is a well-known property, when 
discussing how gravitational waves act on free particles all textbooks present a contradictory result. In fact, as discussed in Section~\ref{standardEFP}, when a first-order gravitational wave passes through orthogonally separated particles, it makes them to oscillate orthogonally around the original position, as depicted in Fig.~1. According to this result, however, the particles are clearly seen to oscillate with the same frequency of the gravitational wave, which is in clear contradiction with the tidal origin of the gravitational waves.

\subsection{Effects on free particles: straight approach}
\label{EffFreePart}

Let us consider again the geodesic deviation equation (\ref{gde2bis}):
\be
\frac{d^2 \xi_{\mbox{\tiny{(1)}}}^\alpha}{ds_{\mbox{\tiny{(0)}}}^2} + \partial_0 \Gammabol_{{\mbox{\tiny{(1)}}}}^\alpha{}_{\beta 0} \; \xi_{\mbox{\tiny{(0)}}}^\beta = \left( \partial_0 \Gammabol_{{\mbox{\tiny{(1)}}}}^\alpha{}_{\beta 0} -
\partial_\beta \Gammabol_{{\mbox{\tiny{(1)}}}}^\alpha{}_{0 0} \right) \xi_{\mbox{\tiny{(0)}}}^\beta.
\label{gde2bisagain}
\ee
According to the standard procedure described in Section~\ref{standardEFP}, by choosing a locally inertial coordinate system, one eliminates the connection term from the left-hand side of Eq.~(\ref{gde2bisagain}). This procedure, however, is mathematically unacceptable. The reason is that, {\em since the wave equation for the metric perturbation $h^{\,\gamma}_{\mt \rho}$ was solved in transverse-traceless coordinates, and considering that ultimately one is going to use that solution in the geodesic deviation equation~(\ref{gde2bisagain}) to obtain the effects of gravitational waves on free particles, mathematical consistency does require that this equation be considered in the very same transverse-traceless coordinates. The use of locally inertial coordinates to get rid of the connection term on left-hand side of Eq.~(\ref{gde2bisagain}) is unacceptable from both mathematical and physical points of view.}

In addition to this question, there is also a conceptual issue. According to the standard approach, when a first-order gravitational wave passes through orthogonally separated particles, it makes them to oscillate orthogonally around the original position. The question then arises: how a strictly attractive interaction like gravitation could give rise to {\em orthogonal oscillations around the original position}? This orthogonal oscillation can be easily understood in the electromagnetic case, where the Lorentz force is either attractive or repulsive depending on the sign of the field component. However, in the {\em strictly attractive case of gravitation}, it is not clear at all how such orthogonal particles oscillation could be possible.

If instead of the long and winding procedure described in Section~\ref{standardEFP}, one accepts the equations in the form they show up from the perturbation analysis, a much more natural and straight approach emerges. To begin with, since the wave equation for $h^{\,\gamma}_{\mt \rho}$ was solved in transverse-traceless coordinates, the geodesic deviation equation~(\ref{gde2bisagain}) must then be considered in the same coordinates. Taking into account that in these coordinates the components $h^{\,\gamma}_{\mt 0}$ vanish, the last term on the right-hand side of Eq.~(\ref{gde2bisagain}) vanishes, and we get 
\be
\frac{d^2 \xi_{\mbox{\tiny{(1)}}}^\alpha}{ds_{\mbox{\tiny{(0)}}}^2} + \partial_0 \Gammabol_{{\mbox{\tiny{(1)}}}}^\alpha{}_{\beta 0} \; \xi_{\mbox{\tiny{(0)}}}^\beta = \partial_0 \Gammabol_{{\mbox{\tiny{(1)}}}}^\alpha{}_{\beta 0} \, \xi_{\mbox{\tiny{(0)}}}^\beta.
\ee
We can then safely cancel out the connection-term $\partial_0 \Gammabol_{{\mbox{\tiny{(1)}}}}^\alpha{}_{\beta 0} \, \xi_{\mbox{\tiny{(0)}}}^\beta$ appearing on both sides of this equation, which reduces to 
\be
\ddot \xi_{\mbox{\tiny{(1)}}}^\alpha = 0.
\label{gde4b}
\ee
This equation says that linear gravitational waves do not produce any effect on free particles \cite{waves}, a result consistent with the fact that linear gravitational waves are unable to transport energy and momentum. The natural way to proceed is then to go to the next order. 

\section{Second-order gravitational waves theory}
\label{5}

\subsection{Second-order gravitational waves}

At the second order, the source energy-momentum tensor is found to be conserved in the covariant sense,
\be
D T^a_\my \equiv d T^a_\my + \Gamma^a_{\mt b} \wedge T_\mt^b = 0,
\label{CoCoLaw}
\ee
with $\Gamma^a_{\mt b}$ the first-order spin connection. As is well-known, it is not a true conservation law, but just an identity, called Noether identity, governing the exchange of energy and momentum between the source and gravitation~\cite{kopo}. This means that, differently from what happens at first order, at the second order a mechanical system can lose energy in the form of gravitational waves. The amount of energy and momentum released is that correctly predicted by the quadrupole radiation formula.
At this order, instead of the Maxwell type field equation (\ref{dH1}), the gravitational field equation acquires the Yang-Mills form
\begin{equation}
dH^\my_a - \kappa \, t^\my_a = \kappa \, T^\my_a,
\label{dH2}
\end{equation}
with $t^\my_a$ the second-order gravitational energy-momentum pseudotensor, which is quadratic in the first-order field variable. Similarly to a gluon field, which is able to transport color charge, the plane-wave solution of the sourceless version of the field equation~(\ref{dH2}) might be able to transport energy and momentum, with the amount of  energy and momentum transported given by the pseudotensor $t^\my_a$. 
From Eq.~(\ref{HarmoCoor}), the second-order harmonic coordinate condition is found to be
\be
\eta^{\mu \nu} \, \Gamma_\my^\lambda{}_{\mu \nu} -
h_\mt^{\mu \nu} \, \Gamma_\mt^\lambda{}_{\mu \nu} = 0.
\ee
Using the first-order solution~(\ref{pw}), it assumes the form
\be
\partial_\mu h_\my^{\,\mu}{}_\lambda - {\textstyle{\frac{1}{2}}} \partial_\lambda h_\my =
- {\textstyle{\frac{i}{2}}} \Phi_\my k_\lambda \exp[{i 2 k_\rho x^\rho}]
\label{hc2}
\ee
where $h_\my = h_\my^{\,\alpha}{}_\alpha$ and
\be
\Phi_\my = A^{\,\rho}_{\mt \sigma} A^{\,\sigma}_{\mt \rho}.
\label{fi2}
\ee
In these coordinates, the second-order wave equation is found to be
\begin{equation}
\Box\,  h_{\my}^{\mu}{}_{\nu} = \onehalf \, {\Phi_\my} \, k^\mu k_\nu
\exp[{i 2 k_\rho x^\rho}].
\label{ffee}
\end{equation}
A monochromatic wave solution to this equation is \cite{gw2}
\be
h_{\my}^{\, \mu}{}_{\nu} = \left(A_{\my}^{\,\mu}{}_{\nu} + i B_{\my}^{\,\mu}{}_{\nu} \right)
\exp[{i 2 k_\rho x^\rho}]
\label{PhysGraWave}
\ee
where
\be
A_{\my}^{\,\mu}{}_{\nu} = \frac{3}{16} \, \Phi_\my \, \delta^\mu_\nu
\qquad \mbox{and} \qquad 
B_{\my}^{\,\mu}{}_{\nu} = - \, \frac{\Phi_\my}{8} \, 
\frac{K_\theta x^\theta}{K_\sigma k^\sigma} \, k^\mu k_\nu
\label{B2}
\ee
are the second-order amplitudes, with $K_\theta$ an arbitrary wave number four-vector. It is important to remark that, owing to the quadratic nonlinearity of the wave equation, the second-order gravitational wave naturally emerges propagating with a frequency that is twice the frequency of the source. This is in agreement with the quadrupole radiation formula, and also with the tidal origin of the waves. The amplitude of the first part of the solution satisfies
\be
A_\my^{\,\mu}{}_{\mu} \equiv A_\my = \textstyle{\frac{3}{4}} \, \Phi_\my \qquad \mbox{and} \qquad
k_\mu A_{\my}^{\,\mu}{}_{\nu} = \frac{1}{4} \, k_\nu A_\my
\label{A2vinculo}
\ee
whilst the amplitude of the second part of the solution satisfies
\be
B_\my^{\,\mu}{}_{\mu} \equiv B_\my = 0 \qquad \mbox{and} \qquad k_\mu B_{\my}^{\,\mu}{}_{\nu} = 0.
\label{ttPart}
\ee

Let us consider again a laboratory proper frame~---~endowed with a Cartesian coordinate system~---~from which the wave will be observed. If the wave is traveling in the $z$ direction of the Cartesian system, the wave vector is that given by Eq.~(\ref{kzdir}). In this case, the amplitude $A_{\my}^{\mu}{}_{\nu}$ turns out to be
\begin{gather}
\left( A_{\my}^{\,\mu}{}_{\nu} \right) = \frac{3 \Phi_\my}{16} \left(
\begin{matrix}
 1 & 0 & 0 & 0 \\
 0 &1 & 0 & 0 \\
 0 & 0 & 1 & 0 \\
 0 & 0 & 0 & 1
\end{matrix}
\right)
\label{A}
\end{gather}
from where we see that this part of the solution has just one physical component:
\be
A^{\;t}_{\my t} = A^{\;x}_{\my x} = A^{\;y}_{\my y} = A^{\;z}_{\my z} \equiv \textstyle{\frac{3}{16}} \, \Phi_\my.
\ee
On the other hand, choosing the arbitrary wave vector $K_\rho$ in such a way that $K_0 = K_1 = K_2 = 0$,\footnote{This choice corresponds to assuming that the wave amplitude $B_{\my}^{\mu}{}_{\nu}$ depends on the distance $z$ from the source, but not on the time.} the amplitude $B_{\my}^{\mu}{}_{\nu}$ turns out to be
\begin{gather}
\left( B_{\my}^{\,\mu}{}_{\nu} \right) = - \, \frac{\Phi_\my z \, \omega}{8 c} \left(
\begin{matrix}
 1 & 0 & 0 & -1 \\
 0 & 0 & 0 & 0 \\
 0 & 0 & 0 & 0 \\
 1 & 0 & 0 & -1
\end{matrix}
\right).
\label{B2z}
\end{gather}
From this expression we see that this part of the solution has two physical components:
\be
B^{0}_{\my 0} = - B^{z}_{\my z} \qquad \mbox{and} \qquad B^{0}_{\my z} = -
B_{\my z}{}^{0}.
\label{PhysCom0}
\ee
It is important to notice that the amplitude $B_{\my}^{\,\mu}{}_{\nu}$ depends on the frequency of the wave, a typical property of nonlinear waves.

It is interesting to observe that the above second-order solution does not represent corrections to the first-order wave, but rather shows up as a completely different solution. This picture bears some resemblance to the case of nonlinear surface waves in shallow water, where solitary waves can be obtained from the Navier-Stokes equation (for an inviscid fluid) through a perturbation scheme \cite{dodd}. At the first order, one obtains a linear wave-equation whose solution determines the dispersion relation of the system, {\em not the physical wave}. At the second order, the first-order solution appears multiplied by itself, giving rise to a nonlinear evolution equation---the so-called Korteweg-de Vries equation. The solitary wave, which is the physical wave observed in nature, is then obtained as a solution to this nonlinear equation. It is important to observe that the physical, second-order solution does not represent a correction to the first-order solution. Rather, it is a completely different (nonlinear) wave. This means that even for very small wave amplitudes, a solitary wave can never be approximately described by a linear equation. As a nonlinear phenomenon, gravitational waves seem to share the same properties.

\subsection{Second-order effects on free particles}

The amplitude of a second-order gravitational wave is in general assumed to fall off with $1/z^2$, with $z$ the distance from the source. Due to the large distances from the sources, second-order effects are usually considered to be neglectful. This is the case of the amplitude $A_{\my}^{\,\mu}{}_{\nu}$ given by (\ref{A}). However, owing to the intricacies of nonlinear differential equations, the amplitude $B_\my^{\mu\nu}$ turns out to present an additional linear dependency on the distance $z$ from the source, which makes it to scale with distance as $1/z$. Contrary to the usual belief, therefore, second-order effects can be relevant for the phenomenology of gravitational waves. From now on, for the purpose of obtaining the effects on free particles, we are going to neglect the $A$-part of solution (\ref{PhysGraWave}), which amounts to assume that the second-order solution is given by
\be
h_{\my}^{\, \mu}{}_{\nu} = i B_{\my}^{\,\mu}{}_{\nu} \exp[{i 2 k_\rho x^\rho}].
\label{PhysGraWave2}
\ee

At second order, the geodesic deviation equation (\ref{gde0}) assumes the form
\be
\frac{d^2 \xi_{\mbox{\tiny{(2)}}}^\alpha}{ds_{\mbox{\tiny{(0)}}}^2} + \Gammabol_{{\mbox{\tiny{(1)}}}}^\alpha{}_{\gamma 0} \; \Gammabol_{{\mbox{\tiny{(1)}}}}^\gamma{}_{\beta 0} \; \xi_{\mbox{\tiny{(0)}}}^\beta + \partial_0 \Gammabol_{{\mbox{\tiny{(2)}}}}^\alpha{}_{\beta 0} \; \xi_{\mbox{\tiny{(0)}}}^\beta =
\Rbol_{\mbox{\tiny{(2)}}}^\alpha{}_{\mu \nu \beta} \, U^\mu_\mz \, U^\nu_\mz \, \xi_{\mbox{\tiny{(0)}}}^\beta
\label{gde2A}
\ee
with
\be
R^{\alpha}_{\my 00 \beta} = \partial_0 \Gamma^\alpha_{\my 0 \beta} -
\partial_\beta \Gamma^\alpha_{\my 0 0} +
\Gamma^\alpha_{\mt 0 \gamma} \Gamma^\gamma_{\mt 0 \beta} -
\Gamma^\alpha_{\mt \beta \gamma} \Gamma^\gamma_{\mt 0 0}
\ee
the second-order Riemann tensor. In the proper frame, where $U_{\mbox{\tiny{(0)}}}^\mu = \delta^\mu_0$, and specialising to transverse-traceless coordinates, the above geodesic deviation equation reduces to
\be
\frac{d^2 \xi_{\mbox{\tiny{(2)}}}^\alpha}{ds_{\mbox{\tiny{(0)}}}^2} = \big(
\onehalf \partial_\beta \partial^\alpha h_{\my 00} -
\partial_\beta \partial_0 h^\alpha_{\my 0} \big) \xi_{\mbox{\tiny{(0)}}}^\beta.
\label{gde4}
\ee
Suppose now two particles separated initially in the $x$ direction by a distance $\xi_{\mbox{\tiny{(0)}}}^x$, that is,
\be
\xi_{\mbox{\tiny{(0)}}}^\beta = (0, \xi_{\mbox{\tiny{(0)}}}^x, 0, 0).
\ee
Considering a gravitational wave traveling in the $z$ direction of the laboratory Cartesian coordinates, it is then easy to verify that the resulting equations of motion are
\be
\ddot \xi_{\mbox{\tiny{(2)}}}^x =
\ddot \xi_{\mbox{\tiny{(2)}}}^y =
\ddot \xi_{\mbox{\tiny{(2)}}}^z = 0,
\label{gde5}
\ee
where we have used that in the proper frame $s_{\mbox{\tiny{(0)}}} = ct$. The same result is obtained for two particles separated initially in the $y$ direction. We consider now two particles separated initially in the $z$ direction by a distance $\xi_{\mbox{\tiny{(0)}}}^z$, that is,
\be
\xi_{\mbox{\tiny{(0)}}}^\beta = (0, 0, 0, \xi_{\mbox{\tiny{(0)}}}^z).
\ee
In this case, the geodesic deviation equation (\ref{gde4}) yields
\be
\ddot \xi_{\mbox{\tiny{(2)}}}^x =
\ddot \xi_{\mbox{\tiny{(2)}}}^y = 0
\label{gde7}
\ee
and
\be
\ddot \xi_{\mbox{\tiny{(2)}}}^z = c^2 \left(\partial_0 \partial_z h_{\my z0} - \onehalf \partial_z \partial_z h_{\my 00} \right) \xi_{\mbox{\tiny{(0)}}}^z.
\label{gde8}
\ee
Using the solution (\ref{PhysGraWave2}), this equation assumes the form
\be
\ddot \xi_{\mbox{\tiny{(2)}}}^z = - {\textstyle{\frac{1}{4}}} \,
\xi_{\mbox{\tiny{(0)}}}^z \Phi_\my \, \frac{z \, \omega^3}{c} \,
\sin[2\omega(t - z/c)].
\ee
Although the coefficient depends on distance $z$, for practical purposes it can be assumed to be constant in the region of the experience. We then write
\be
\ddot \xi_{\mbox{\tiny{(2)}}}^z = - \textstyle{\frac{1}{4}} \,
\xi_{\mbox{\tiny{(0)}}}^z \tilde{\Phi}_\my \, \omega^2
\sin[2(\omega t - z/{\lambdabar})]
\label{gde9}
\ee
where
\be
\tilde{\Phi}_\my = \Phi_\my \frac{z}{\lambdabar}
\ee
is now assumed to be constant, with $\lambdabar = c/\omega$ the reduced wavelength.
With this assumption, the origin of the coordinate $z$ turns out to be completely arbitrary. We can then choose one of the particles to be at $z=0$, in which case $z$ will represent the position of the second particle. Assuming that initially the particles are at rest, that is, for $t = 0$ we have $\xi_{\mbox{\tiny{(2)}}}^z = \xi_{\mbox{\tiny{(0)}}}^z$, the solution of the geodesic deviation equation~(\ref{gde9}) is found to be
\be
\xi_{\mbox{\tiny{(2)}}}^z = \xi_{\mbox{\tiny{(0)}}}^z + \textstyle{\frac{1}{16}} \,
\xi_{\mbox{\tiny{(0)}}}^z \tilde{\Phi}_\my \Big[ \sin[2(\omega t - z/{\lambdabar})] +
\sin[2 z/{\lambdabar}] \Big].
\label{sol4}
\ee
The only role of the last term on the right-hand side is to make the solution comply with the above initial condition.

When a gravitational wave passes through two particles separated by a distance $\xi_{\mbox{\tiny{(0)}}}^z$ {\em along the direction of propagation of the wave}, both particles begin moving towards the source due to the attraction of gravitation. In addition, owing to the inhomogeneity of the field component $h^{\,0}_{\my z} = h_{\my z}{}^{0}$, sometimes the first particle will move faster than the second, sometimes the second will move faster than the first, in such a way that the distance between the two particles oscillates as they move. Notice that this kind of oscillation does not require that the gravitational interaction changes sign---that is, become repulsive---during the process. It is a genuine tidal oscillation, which is fully consistent with the strictly attractive character of the gravitational interaction. Notice also that, differently from an ordinary longitudinal oscillation around the equilibrium position, the combination {\em longitudinal motion} plus {\em oscillation} does have a quadrupole nature. This combined effect is the only kind of oscillation that can be produced by tidal effects, and is the signal to be looked for when searching for gravitational waves.

\section{Final remarks}

In this review, the existence of several obscure points and inconsistencies in the standard approach to the gravitational waves theory has been pointed out. A natural way to circumvent these problems is to resignedly accept as correct all results emerging from the first-order expansion of Einstein equation, which in turn amounts to accept that {\em gravitational waves cannot be described by a linear equation, even approximately}. One should then go to the second order, where a sound and consistent gravitational wave theory shows up. The second-order gravitational wave is found to be {\em longitudinal}, in agreement with the strictly attractive character of gravitation. Furthermore, it is found to propagate with a frequency that is twice the source frequency, in agreement with the quadrupole radiation formula, as well as with the tidal origin of gravitational waves. Finally, in contrast to the usual belief, according to which second-order effects should scale as $1/z^2$, with $z$ the distance from the source, the longitudinal second-order wave is find to fall off as $1/z$.

It is clear by now that none of the existing antennas has succeeded in detecting any sign of gravitational waves. Of course, it is possible that the detectors did not meet the necessary sensibility to detect them, or that the magnitude of the gravitational waves when reaching a detector on Earth is smaller than originally predicted. However, it is also possible that a faulty approach has led all detectors to look for the wrong sign. The analysis presented in these notes suggests that this possibility should not be neglected.

\section*{Acknowledgments}
The author would like to thank R. Aldrovandi, R. da Rocha and K. H. Vu for useful discussions. He would like to thank also FAPESP, CNPq and CAPES for partial financial support.


\begin{thebibliography}{99}
\bibitem{all}
This argument is present in any textbook or review article on gravitational waves.
\end{thebibliography}
\end{document}